\documentclass[11pt]{article}

\newsavebox{\foobox}
\newcommand{\slantbox}[2][0]{\mbox{%
        \sbox{\foobox}{#2}%
        \hskip\wd\foobox
        \pdfsave
        \pdfsetmatrix{1 0 #1 1}%
        \llap{\usebox{\foobox}}%
        \pdfrestore
}}
\newcommand\unslant[2][-.25]{\slantbox[#1]{$#2$}}

\newcommand{\mpi}{\text{\unslant[-.18]\pi}}

\usepackage[left=2cm, right=2cm, top=2.5cm, bottom=2.5cm]{geometry}
\usepackage[x11names]{xcolor}
\usepackage{fancyhdr, amssymb, cancel, amsmath, graphicx, pgfplots, tikz}
\usepackage{isomath}

\usetikzlibrary{shadows}

\newcommand{\stylecolor}{violet}

\usepackage[labelfont={bf,sf, color=\stylecolor}, margin={1.5cm,0cm}]{caption}

\usepackage[colorlinks=true, urlcolor=\stylecolor!70!white, linkcolor=\stylecolor, citecolor=\stylecolor!70!white, hyperindex=true, linktocpage=true]{hyperref}

\usepackage[explicit]{titlesec}

\newcommand*\sectionlabel{}
\titleformat{\section}
  {\gdef\sectionlabel{}
   \Large\bfseries\scshape}
  {\gdef\sectionlabel{\thesection }}{0pt}
  {\begin{tikzpicture}[remember picture,overlay]
       \end{tikzpicture}
  }
\titlespacing*{\section}{0pt}{0pt}{0pt}

\newcommand*\subsectionlabel{}
\titleformat{\subsection}
  {\gdef\subsectionlabel{}
   \large\bfseries\scshape}
  {\gdef\subsectionlabel{\thesubsection  }}{0pt}
  {\begin{tikzpicture}[remember picture]
    	\draw (-0.15, 0) node[left] {\color{\stylecolor} \textsf{\subsectionlabel}};
	\draw (0.15, 0) node[right] {\color{\stylecolor} \textsf{#1}};
	\fill[color=\stylecolor] (-0.05, -0.23) rectangle (0.05, 0.23);
       \end{tikzpicture}
  }
\titlespacing*{\subsection}{-4pt}{10pt}{0pt}

\newcommand*\subsubsectionlabel{}
\titleformat{\subsubsection}
  {\gdef\subsubsectionlabel{}
   \bfseries\scshape}
  {\gdef\subsubsectionlabel{\thesubsubsection.\ \  }}{0pt}
  {\begin{tikzpicture}[remember picture]
    	\draw (0, 0) node[left] {\color{\stylecolor} \textsf{\subsubsectionlabel}};
	\draw (0, 0) node[right] {\color{\stylecolor} \textsf{#1}};
       \end{tikzpicture}
  }
\titlespacing*{\subsubsection}{-4pt}{7pt}{0pt}

\pgfplotsset{every axis legend/.append style={at={(1.02,1)},anchor=north west}}

\begin{document}

\allowdisplaybreaks

\pagestyle{fancy}
\renewcommand{\headrulewidth}{0pt}
\fancyhead{}

\fancyfoot{}
\fancyfoot[C] {\textsf{\textbf{\thepage}}}

\begin{equation*}
\begin{tikzpicture}
\draw (\textwidth, 0) node[text width = \textwidth, right] {\color{white} easter egg};
\end{tikzpicture}
\end{equation*}

\begin{equation*}
\begin{tikzpicture}
\draw (0.5\textwidth, -3) node[text width = \textwidth] {\huge  \textsf{\textbf{Origin of the Drude peak and of zero sound  in probe  \\ \vspace{0.07 in} brane holography}} };
\end{tikzpicture}
\end{equation*}
\begin{equation*}
\begin{tikzpicture}
\draw (0.5\textwidth, 0.1) node[text width=\textwidth] {\large \color{black}  \textsf{Chi-Fang Chen and Andrew Lucas}};
\draw (0.5\textwidth, -0.5) node[text width=\textwidth] {\small\textsf{Department of Physics, Stanford University, Stanford, CA 94305, USA}};
\end{tikzpicture}
\end{equation*}
\begin{equation*}
\begin{tikzpicture}
\draw (0, -13.1) node[right, text width=0.5\paperwidth] {\texttt{chifangc@stanford.edu, ajlucas@stanford.edu}};
\draw (\textwidth, -13.1) node[left] {\textsf{\today}};
\end{tikzpicture}
\end{equation*}
\begin{equation*}
\begin{tikzpicture}
\draw[very thick, color=\stylecolor] (0.0\textwidth, -5.75) -- (0.99\textwidth, -5.75);
\draw (0.12\textwidth, -6.25) node[left] {\color{\stylecolor}  \textsf{\textbf{Abstract:}}};
\draw (0.53\textwidth, -6) node[below, text width=0.8\textwidth, text justified] {\small At zero temperature, the charge current operator appears to be conserved, within linear response, in certain holographic probe brane models of strange metals.  At small but finite temperature, we analytically show that the weak non-conservation of this current leads to both a collective ``zero sound" mode and a Drude peak in the electrical conductivity.  This simultaneously resolves two outstanding puzzles about probe brane theories.  The nonlinear dynamics of the current operator itself appears qualitatively different.};
\end{tikzpicture}
\end{equation*}

\tableofcontents

\begin{equation*}
\begin{tikzpicture}
\draw[very thick, color=\stylecolor] (0.0\textwidth, -5.75) -- (0.99\textwidth, -5.75);
\end{tikzpicture}
\end{equation*}

\titleformat{\section}
  {\gdef\sectionlabel{}
   \Large\bfseries\scshape}
  {\gdef\sectionlabel{\thesection }}{0pt}
  {\begin{tikzpicture}[remember picture]
	\draw (0.2, 0) node[right] {\color{\stylecolor} \textsf{#1}};
	\draw (0.0, 0) node[left, fill=\stylecolor,minimum height=0.27in, minimum width=0.27in] {\color{white} \textsf{\sectionlabel}};
       \end{tikzpicture}
  }
\titlespacing*{\section}{0pt}{20pt}{5pt}

\section{Introduction}
One of the earliest applications of the AdS/CFT correspondence to condensed matter physics was to study holographic ``probe branes" at finite density \cite{kobayashi, karch}.    The holographic dual of such models is (in the simplest cases) widely believed to be  $\mathcal{N}=2$ supersymmetric fundamental matter (analogous to quarks), localized on a defect within the $\mathcal{N}=4$ supersymmetric Yang-Mills plasma  \cite{karch2}.    Because electrical transport in strongly interacting quantum systems remains a challenging problem in condensed matter,  much of the work on these probe brane models focuses on the transport of the conserved U(1) baryon number.   Unfortunately, the probe limit, where the background plasma is unaffected by the dynamics of the baryon matter, leads to certain simplifying features of transport that are absent in more ``realistic" holographic models for strange metals \cite{koenraadbook, lucasreview}.

 It is still important to understand the transport properties of the probe brane models, however;  they remain rare examples of solvable interacting quantum systems in higher dimensions.    And at low temperature, the behavior of probe branes has remained rather mysterious for almost a decade.    Firstly, at low temperatures one often finds a collective, propagating ``sound mode" \cite{karch08, parnachev, parnachev2, wu, davison11}.   This cannot be ordinary sound, as the energy-momentum tensor is dominated by the `decoupled' $\mathcal{N}=4$ plasma;  it also cannot be (superfluid) second sound as the U(1) symmetry is not broken.   From this line of thought, \cite{karch08} subsequently concluded that this propagating mode was analogous to zero sound:  the sloshing of the Fermi surface in a Fermi liquid at low temperature \cite{pines}.   While there is no strong evidence for a well-defined baryonic Fermi surface, save for finite momentum spectral weight \cite{anantua}, we will follow the literature and call this propagating mode zero sound.   Furthermore, when the zero sound waves are present, the low frequency electrical conductivity $\sigma(\omega)$ has an apparent Drude peak at low temperature \cite{hartnoll09}.     Such a Drude peak would normally be associated with approximate conservation of momentum \cite{hartnollhofman, lucasreview}, but as we have already mentioned, that cannot be the case in probe brane models.

We demonstrate below that within linear response in probe brane holography, the charge current operator itself appears to be conserved at zero temperature when the dynamical critical exponent $z$ of the background plasma obeys $z<2$.    This emergent conservation law, and the resulting ``hydrodynamics", is responsible for the Drude singularity in the electrical conductivity, as well as the propagation of the zero sound mode.     At small but finite temperature, the charge current decays at a rate $\sim T^{2/z}$;  this decay is responsible for both the breakdown of zero sound modes as well as the broadening of the Drude peak.   We also emphasize that this low temperature hydrodynamics is distinct from the high temperature hydrodynamics of probe branes, which is conventional, and describes a single diffusive mode for charge.

This mechanism is analogous to the behavior of electrical conductivity \cite{lucasreview, hartnollhofman, lucasMM} and ordinary sound waves \cite{lucasplasma} in a normal fluid with weak momentum relaxation;  such similarity was qualitatively observed before  \cite{gouteraux2}.     Here, of course, the conserved momentum is replaced by the charge current operator itself.   Unlike the hydrodynamics of ordinary fluids, we find that the nonlinear hydrodynamics of the current operator is often ill-posed:  the gradient expansion always fails at low enough temperatures.   Thus, as we resolve the mysteries of the zero sound modes and the Drude conductivity which arise within linear response, our work also calls into question the ultimate fate of the zero sound mode, and of electrical transport more broadly, at the nonlinear level.    

\section{Probe Branes at Finite Density}
Consider a large $N$ quantum field theory (QFT) in $d$ spatial dimensions with a conserved U(1) current $J^\mu$.  If this QFT has a holographic dual,  then $J^\mu$ is dual to a bulk gauge field $A^a$ ($ab\cdots$ denote bulk indices;  $\mu\nu\cdots$ denote boundary indices).   For probe brane models, the generating functional for correlators of $J^\mu$ is simply the exponential of the Dirac-Born-Infeld (DBI) action for $A^a$:  \begin{equation}
S =  \mathcal{K} \int \mathrm{d}^{d+2}x \sqrt{-\det\left(g_{ab} + 2\mpi \alpha^\prime F_{ab}\right)}  \label{eq:S}
\end{equation}
where $F = \mathrm{d}A$, $g_{ab}$ is the bulk spacetime metric and $\alpha^\prime$ is the string tension, a dimensionful constant arising from the  string theory interpretation of the holographic dual \cite{lucasreview}.   The constant $\mathcal{K}$ is related to the brane tension and the volumes of any compact wrapped spaces, and is unimportant for us.    In this paper, we consider background metrics of ``Lifshitz" form \cite{kachru}: \begin{equation}
\mathrm{d}s^2 = \frac{L^2}{r^2} \left[\frac{\mathrm{d}r^2}{f(r)} - \frac{f(r)}{r^{2z-2}} \mathrm{d}t^2 + \mathrm{d}x^i \mathrm{d}x_i\right]
\end{equation}
The parameter $z$ is called the dynamical critical exponent, and is related to the relative scaling of time and space in the dual critical theory; we restrict to theories with $z\ge 1$ \cite{lucasreview}.   The function $f(r)$ encodes a finite temperature $T$, and is given by \begin{equation}
f(r) = 1-\left(\frac{r}{r_{\mathrm{h}}}\right)^{d+z}
\end{equation}
with \begin{equation}
r_{\mathrm{h}}^z = \frac{d+z}{4\mpi T}.  \label{eq:Tz1}
\end{equation}
   In the probe brane limit, $g_{ab}$ is a fixed, non-dynamical background field.  

If we are interested in studying matter at finite density, then we must look for saddle points of the action (\ref{eq:S}).   These are solutions to  \begin{equation}
\partial_a \left( \sqrt{-\det(g+2\mpi \alpha^\prime F)} (g+2\mpi \alpha^\prime F)^{[ab]} \right) = 0.  \label{eq:eom}
\end{equation}
In the above equation, $(g+2\mpi \alpha^\prime F)^{ab}$ refers to components of the matrix inverse of $g_{ab} + 2\mpi \alpha^\prime F_{ab}$.    


\section{Hydrodynamics of the Conserved Current}
Now, let us look for a solution to (\ref{eq:eom}) in which $\rho$ is a slowly varying function of the boundary theory coordinates $x^\mu$.  In other words, we perform a gradient expansion, keeping track of terms only to lowest order in the number of $x^\mu$-derivatives ($r$-derivatives, denoted with $\prime$, will not be treated as perturbatively small). Working in radial gauge $A_r=0$, we find that the $\mu$-components of (\ref{eq:eom}) give 
\begin{subequations}\label{eq:gradex}\begin{align}
 \left( \frac{1}{\mathcal{L}} \left(\frac{L}{r}\right)^{2d}  A_t^\prime \right)^\prime &=  \mathcal{O}\left(\partial_\mu^2 \right), \label{eq:gradexa} \\
  \left( \frac{1}{\mathcal{L}} \left(\frac{L}{r}\right)^{2d} \frac{f}{r^{2z-2}} A_i^\prime \right)^\prime &=  \mathcal{O}\left(\partial_\mu^2 \right),  \label{eq:gradexb}
\end{align}\end{subequations}
where \begin{equation}
\mathcal{L} = \frac{L^{d+2}}{r^{d+1+z}} \sqrt{1  - \frac{(2\mpi\alpha^\prime)^2}{L^4} \left[ r^{2+2z} A_t^{\prime 2} - fr^4 A_x^{\prime 2}\right] } + \mathcal{O}\left(\partial_\mu^2\right)  \label{eq:L3}
\end{equation}
Note that either the presence of finite temperature $T$, or $z>1$, breaks the symmetry between $t$ and $x^i$.   These equations can be exactly solved by 
\begin{subequations}\label{eq:Asoln}\begin{align}
A_t^\prime &= - \rho(x^\mu)  \frac{r^{2d}\mathcal{L}}{L^{d+2}}, \\
A_i^\prime &= J_i(x^\mu)  \frac{r^{2d}\mathcal{L}}{L^{d+2}} \frac{r^{2z-2}}{f},
\end{align}\end{subequations}
where we can also write\begin{equation}
\mathcal{L} = \frac{L^{d+2}}{r^{d+1+z}} \dfrac{1}{\displaystyle \sqrt{1+ C^2  \left(\rho^2 - \dfrac{r^{2z-2}}{f} (J^i)^2\right) r^{2d}  }}
\end{equation}
with \begin{equation}
C^2 = \dfrac{(2\mpi\alpha^\prime)^2}{L^4}.
\end{equation}
Here $\rho(x^\mu)$ and $J^i(x^\mu)$ can be interpreted as the charge density and charge current in the dual theory through a rescaling of $\mathcal{K}$; we will assume this henceforth.

So far, this solution is ``exact".   However, it is clear that if $z>1$ or $T>0$, the nonlinear solutions with any finite $J^i$ are \emph{not well-posed}.    This will lead to the breakdown of hydrodynamics at nonlinear order, but we defer this discussion to Section \ref{sec:nl}.   At $T=0$, and $z=1$, we recover the ``boost" solutions of \cite{thompson}, whose existence is demanded by Lorentz covariance.

\subsection{Linear Response}\label{sec:LR}
To proceed farther, let us assume that $J^i$ is infinitesimally small, and only consider first order terms in $J^i$.  This is a rather artificial limit to ensure that the solution (\ref{eq:Asoln}) exists.   However, this linear response regime is precisely where both zero sound and the Drude peak are observed, and so a careful understanding of this regime is sufficient to understand these phenomena.   Thus, we proceed.    We will see that within this linear response limit, we should treat $J^i$ and $\partial_\mu \rho$ as infinitesimal quantities, and so they need only be kept to linear order.   


Let us first perform the integral over $r$ in $A_t^\prime$:  \begin{align}
A_t(r) = A^0_t - \rho \int\limits_0^r \mathrm{d}s  \frac{s^{d-1-z}}{\sqrt{1+C^2 \rho^2 s^{2d}}} = A^0_t - C^{-1+z/d} \rho^{z/d} \mathcal{F}_1 \left((C\rho)^{1/d}r\right)  \label{eq:At}
\end{align}
where $A^0_t$ is a constant of integration,  physically dual to a background gauge field coupled to the current operator $J^\mu$ in the dual theory, and \begin{equation}
\mathcal{F}_1(x) \equiv \int\limits_0^x \mathrm{d}y  \frac{y^{d-z-1}}{\sqrt{1+y^{2d}}}.
\end{equation}
 Implicitly, of course, $\rho$ and $J^i$ depend on $x^\mu$, and we will drop the explicit dependence henceforth.  For large $x$, we find the asymptotic expansion \begin{equation}
\mathcal{F}_1(x) = c_1 - \frac{b_1}{x^z} + \cdots
\end{equation}
with positive coefficient \begin{equation}
c_1 = \frac{1}{2\sqrt{\mpi}d} \mathrm{\Gamma}\left(\frac{d-z}{2d}\right)\mathrm{\Gamma}\left(\frac{z}{2d}\right)
\end{equation}
and $b_1>0$;  this will come in handy soon.   Note that $A^0_t$ is not arbitrary, and should be chosen so that $A_t$ vanishes on the horizon \cite{lucasreview}.

We now perform the $r$ integral in $A_i^\prime$: \begin{align}
A_i(r) &= A^0_i + J^i \int\limits_0^r \mathrm{d}s  \frac{s^{d+z-3}}{f(s)\sqrt{1+C^2 \rho^2 s^{2d}}} \notag \\
&= A^0_i + J_i (C\rho)^{(2-z-d)/d}  \mathcal{F}_2 \left((C\rho)^{1/d}r\right) + J_i \int\limits_0^r \frac{\mathrm{d}s}{\sqrt{1+C^2\rho^2 s^{2d}}}   s^{d+z-3} \left(\frac{s}{r_{\mathrm{h}}}\right)^{d+z} \left[1-\left(\frac{s}{r_{\mathrm{h}}}\right)^{d+z} \right]^{-1}  \label{eq:Ax}
\end{align}
where \begin{equation}
\mathcal{F}_2(x) = \int\limits_0^x \mathrm{d}y   \frac{y^{d+z-3}}{\sqrt{1+y^{2d}}}. 
\end{equation}
Depending on the value of $z$,  $\mathcal{F}_2$ has qualitatively different behavior.   For $z<2$, this integral is convergent  and one finds \begin{equation}
\mathcal{F}_2(x) = c_2 - \frac{b_2}{x^{2-z}} + \cdots
\end{equation}
with \begin{equation}
c_2 = \frac{1}{2\sqrt{\mpi}d} \mathrm{\Gamma}\left(\frac{d+z-2}{2d}\right)\mathrm{\Gamma}\left(\frac{2-z}{2d}\right)
\end{equation}
In the limit of low temperatures ($(C\rho)^{1/d} r_{\mathrm{h}} \gg 1$), the second term of (\ref{eq:Ax}) has a logarithmic divergence near the horizon,  and so for $r\approx r_{\mathrm{h}}$ we cannot neglect the second term: \begin{align}
\int\limits_0^r \frac{\mathrm{d}s}{\sqrt{1+C^2\rho^2 s^{2d}}}   s^{d+z-3} \left(\frac{s}{r_{\mathrm{h}}}\right)^{d+z} \left[1-\left(\frac{s}{r_{\mathrm{h}}}\right)^{d+z} \right]^{-1} &\approx  \frac{1}{C\rho}\int\limits_0^r \mathrm{d}s  \frac{s^{d+2z-3}}{r_{\mathrm{h}}^{d+z}-s^{d+z}} + \mathcal{O}\left(\frac{T^{2d/z}}{\rho^2}\right) \notag \\
&\approx \frac{r_{\mathrm{h}}^{z-2}}{C\rho} \log \frac{r_*}{r_{\mathrm{h}}-r} + \text{constant}.  \label{eq:Tlog}
\end{align}  
The coefficient $r_*$ is a constant which is not important.   
For $z\ge 2$, $\mathcal{F}_2$ diverges at large $x$.     The coefficient of the logarithmic term in (\ref{eq:Tlog}) also diverges in this limit.   Thus when $z\ge 2$ our gradient expansion fails, and we will explain what happens in this limit briefly in Section \ref{sec:zerosound}.

Not every $A_t$ given by (\ref{eq:At}) and $A_x$ given by (\ref{eq:Ax}) is a solution of (\ref{eq:eom}).  We now must go to first order in the gradient expansion to find the physically allowed solutions.  Using the $r$-component of (\ref{eq:eom}), we obtain \begin{equation}
0 = -\partial_t \left(\frac{1}{\mathcal{L}} \left(\frac{L}{r}\right)^{2d} A_t^\prime\right) + \partial_i \left(\frac{f}{\mathcal{L} r^{2z-2}} \left(\frac{L}{r}\right)^{2d} A_i^\prime\right).  \label{eq:bulkcurrentcons}
\end{equation}
Using (\ref{eq:Asoln}) we immediately find \begin{equation}
\partial_t \rho + \partial_i J^i = 0,
\end{equation} 
which implies that the expectation value of the current operator $J^\mu$ is conserved in the dual theory.  This is of course a physical requirement, and not surprising.

A more subtle point is that the presence of a logarithmic divergence in $A_i$ at any finite $T$ would lead to a breakdown of our linear response theory.   However, this is an artifact of the approximation (\ref{eq:gradex}).  In fact, we will show that when $r\rightarrow r_{\mathrm{h}}$, the time derivatives in the $x^i$-components of (\ref{eq:gradex}) cannot be neglected.   So we must treat the near horizon region more carefully: in this region we must replace (\ref{eq:gradexb}) with \begin{equation}
  \left( \frac{1}{\mathcal{L}} \left(\frac{L}{r}\right)^{2d} \frac{f}{r^{2z-2}} A_i^\prime \right)^\prime  = \partial_t \left(\frac{1}{\mathcal{L}} \left(\frac{L}{r}\right)^{2d} \frac{\partial_t A_i - \partial_i A_t}{f}\right).  \label{eq:Eeq0}
\end{equation}
It is easiest to proceed by defining a gauge-invariant quantity \begin{equation}
\mathcal{E}_i = \partial_i A_t - \partial_t A_i,
\end{equation} 
which obeys an approximate near-horizon equation of motion \begin{equation}
  \left( \frac{1}{\mathcal{L}} \left(\frac{L}{r}\right)^{2d} \frac{f}{r^{2z-2}} \mathcal{E}_i^\prime \right)^\prime  \approx \partial_t \left(\frac{1}{\mathcal{L}} \left(\frac{L}{r}\right)^{2d} \frac{\partial_t \mathcal{E}_i}{f}\right).  \label{eq:Eeq}
\end{equation}
We have dropped subleading contributions in $f \sim r_{\mathrm{h}}-r$.  In particular, using (\ref{eq:bulkcurrentcons}), we observe that $A_t^\prime$ is generally small compared to $A_x^\prime$ near the horizon, which justifies taking a time derivative of (\ref{eq:Eeq0}) to obtain (\ref{eq:Eeq}).

To linear order in $\mathcal{E}_i$ and $\partial_t \mathcal{E}_i$, we may simplify this equation in the near-horizon limit $r\approx r_{\mathrm{h}}$:
\begin{equation}
r_{\mathrm{h}}^{2-2z}f\left(f\mathcal{E}_i^\prime\right)^\prime \approx \partial_t^2 \mathcal{E}_i .  \label{eq:near}
\end{equation}
Near the horizon, \begin{equation}
f(r) \approx \frac{d+z}{r_{\mathrm{h}}} (r_{\mathrm{h}}-r) + \cdots.  \label{eq:Tz2}
\end{equation}
Defining \begin{equation}
R \equiv  \frac{1}{4\mpi T} \log \frac{r_{\mathrm{h}}}{r_{\mathrm{h}}-r},
\end{equation}
and using (\ref{eq:Tz1}) and (\ref{eq:Tz2}), (\ref{eq:near}) becomes
\begin{equation}
\partial_R^2 \mathcal{E}_i = \partial_t^2 \mathcal{E}_i.
\end{equation}
The solution obeying the physical boundary conditions (falling into the black hole) can be written as a Fourier transform \begin{equation}
\mathcal{E}_i \approx \int\mathrm{d}\omega \; H_i(\omega) \mathrm{e}^{\mathrm{i}\omega(R-t)}.  \label{eq:mathcalA}
\end{equation}
Note that $\mathcal{E}_i$ is completely regular near the horizon.

Following \cite{lucas1501}, let us now consider -- for fixed $R$ -- the limit where $H_i(\omega)$ only has support for vanishingly small frequencies.   In this limit, we may Taylor expand (\ref{eq:mathcalA}) in the near-horizon limit: \begin{equation}
\mathcal{E}_i(r,t,x^i) \approx H_i(t,r,x^i) - \frac{\partial_t H_i(t,r,x^i)}{4\mpi T} \log \frac{r_{\mathrm{h}}}{r_{\mathrm{h}}-r} + \mathcal{O}\left(\partial_t^2\right).  \label{eq:H}
\end{equation}
This formula for $\mathcal{E}_i$ is valid for $ r_{\mathrm{h}} \exp[-4\mpi T/\omega] \lesssim r_{\mathrm{h}}-r \lesssim r_{\mathrm{h}}$.    In the limit where $\omega \rightarrow 0$, the regime of validity of (\ref{eq:At}) and (\ref{eq:Ax}) is $r_{\mathrm{h}} -r \gtrsim \mathrm{O}(\omega)$.\footnote{The precise prefactor can depend on the value of $z$.}   Therefore, we can obtain a non-trivial constraint on the dynamics by demanding that the values of $\mathcal{E}_i$ obtained using (\ref{eq:At}) and (\ref{eq:Ax}) are consistent with (\ref{eq:H}).  In particular, consider the approximation  \begin{equation}
\mathcal{E}_i \approx \partial_i A^0_t - \partial_t A^0_i - \frac{c_1}{(C\rho)^{1-z/d}} \frac{z}{d} \partial_i \rho - \frac{c_2}{(C\rho)^{(d+z-2)/d}}\partial_t J_i  - \frac{r_{\mathrm{h}}^{z-2}}{C\rho}  \partial_t J_i    \log \frac{r_*}{r_{\mathrm{h}}-r} \label{eq:Ei}
\end{equation}
We have dropped all terms which are subleading in powers of $T$ in the above expression, for simplicity.  Comparing (\ref{eq:H}) to (\ref{eq:Ei}) we conclude that \begin{equation}
 \partial_i A^0_t - \partial_t A^0_i - \frac{c_1}{(C\rho)^{1-z/d}} \frac{z}{d} \partial_i \rho - \frac{c_2}{(C\rho)^{(d+z-2)/d}}\partial_t J_i  =4\mpi T \frac{r_{\mathrm{h}}^{z-2}}{C\rho}  J_i   + q_i(x)
\end{equation}
where $q(x)$ is a $t$-independent function.   In physical circumstances, we must have $q_i=0$ -- consider sources which are switched off at $t=-\infty$, when the fluid is at rest.    Then we clearly have $q_i=0$, which will continue to hold for all times, even if we begin to turn on non-trivial $A^0_\mu$.   

Recognizing the source terms as simply the externally applied electric field $E_i$, we conclude that the ideal linearized hydrodynamics on probe branes is \begin{subequations}\label{eq:sound}\begin{align}
\partial_t \rho + \partial_i J^i &= 0, \\
\partial_t J^i + v^2 \partial_i \rho   &= \chi E^i -  \frac{J^i}{\tau},
\end{align}\end{subequations}
where the decay rate of the weakly non-conserved $J^i$ is\begin{equation}
\frac{1}{\tau} = \frac{(4\mpi T)^{2/z}}{(d+z)^{(2-z)/z}c_2 (C\rho)^{(2-z)/d}},
\end{equation}
the current-current susceptibility is \begin{equation}
\chi =  (C\rho)^{(d+z-2)/d},
\end{equation}
and the speed of sound is \begin{equation}
v^2 = \frac{zc_1}{dc_2 (C\rho)^{(2-2z)/d}}
\end{equation}
in agreement with \cite{wu}.   Note that if $z=1$, $c_1=c_2$ and $v^2 = 1/d$ \cite{karch08}.

(\ref{eq:sound}) is our main result.  It demonstrates that the linearized low temperature hydrodynamics in probe brane models is mathematically equivalent to the hydrodynamics of an ordinary fluid with weak momentum relaxation \cite{lucasreview, lucasplasma}.  This proves a ``conjecture" of \cite{lucasreview, gouteraux2}.   It will now be straightforward to see the emergence of the Drude peak and zero sound, and how both are linked to the same emergent conservation law. 

\subsection{The Drude Peak}
  We begin with the Drude peak in the conductivity.  To compute the electrical conductivity, we apply a spatially homogeneous time-dependent electric field $E_i(\omega) \mathrm{e}^{-\mathrm{i}\omega t}$.    Using (\ref{eq:sound}) together with Ohm's law: \begin{equation}
J_i(\omega)  = \sigma(\omega) E_i(\omega)
\end{equation} 
we obtain 
\begin{equation}
\sigma(\omega) = \frac{\chi\tau}{1-\mathrm{i}\omega \tau}.  \label{eq:drude}
\end{equation} 
This functional form is called the Drude peak, and was numerically observed in \cite{hartnoll09}.  The temperature dependence of $\tau$ is consistent with that found in \cite{davison11, hartnoll09}.   (\ref{eq:drude}) is consistent with the predictions of the memory matrix formalism \cite{lucasreview, hartnollhofman, lucasMM}, when the charge current itself is an almost conserved quantity.   

\subsection{Zero Sound}
\label{sec:zerosound}

Next, we turn to the zero sound modes.   Looking for $\rho$ and $J^i$ proportional to $\mathrm{e}^{\mathrm{i}kx - \mathrm{i}\omega t}$ which solve (\ref{eq:sound}) when $E_i = 0$, we immediately find the dispersion relation \begin{equation}
\omega \left(\omega + \frac{\mathrm{i}}{\tau}\right) = v^2k^2.
\end{equation} 
As already advertised, the speed of zero sound agrees with previous analytic results \cite{karch08, wu}, and the finite temperature decay rate is consistent with the $z=1$, $d=3$ numerics of \cite{davison11}.    While our derivation above implicitly assumed that $\omega \ll T$, so that the Taylor expansion (\ref{eq:H}) was justified, we observe that the zero sound speed is not sensitive to whether $T=0$ or $T>0$, as numerically found in \cite{davison11}.

Let us now briefly turn to the fate of zero sound when $z> 2$.   This was described in \cite{wu}, and we repeat the result:  \begin{equation}
\omega = \eta k^{2z/(z+2)} + \cdots
\end{equation}
with $\eta$ a complex-valued coefficient with $\mathrm{Im}(\eta)<0$.   There are two important features of this result.   Firstly, the fact that $\eta$ is complex implies that the zero sound modes are no longer truly long-lived collective excitations; they are strongly damped.  Secondly,  for $z>2$, this dispersion relation is not analytic in $k$, and this is associated directly with the breakdown of the gradient expansion that we observed in Section \ref{sec:LR}.   The real-space equations governing these strongly damped modes will be non-local, in contrast to (\ref{eq:sound}).   When $z=2$, the dispersion relation is of the form $k\sim \omega \sqrt{\log \omega}$, and so is also nonlocal;  furthermore, the zero sound will be a purely dissipative mode as $\omega \rightarrow 0$.

\subsection{Nonlinear Dynamics}
\label{sec:nl}
We now turn to the fate of this hydrodynamics at the nonlinear level.   While we focus on the case $z=1$ for simplicity, our comments are valid for $z>1$ as well.   
We begin by discussing the $T=0$ limit, and analyze the nonlinear corrections to the linearized equations of motion (\ref{eq:gradex}).   (\ref{eq:L3}) is replaced by \begin{equation}
\mathcal{L} = \frac{L^{d+2}}{r^{d+2}} \sqrt{1 + C^2 r^4 A_\mu^\prime A^{\mu \prime} + C^2 r^4 \partial_{[\mu}A_{\nu]} \partial^{[\mu}A^{\nu]}  } + \mathrm{O}(a_\mu^3), \label{eq:L41}
\end{equation}
Let us now consider a small but finite perturbation $a_\mu$ around the background \begin{equation}
\bar A_i = 0, \;\;\;\; \bar A_t =  \int\limits^\infty_r \mathrm{d}r   \frac{\rho r^{d-2}}{\sqrt{1+C^2\rho^2 r^{2d}}};
\end{equation}
thus $A_\mu = \bar{A}_\mu + a_\mu$.
On the background solution, $1- C^2 r^4\bar{A}_t^{\prime 2} \sim r^{-2d}$ at large $r$.  So  following \cite{karch}, analysis of (\ref{eq:eom}) within linear response at $T=0$ yields the following linear differential equation for $a_\mu$ at large $r$: \begin{equation}
\left(r^2 a_\mu^\prime \right)^\prime + \partial^\nu \left(r^2 (\partial_\nu a_\mu - \partial_\mu a_\nu)  \right) \approx 0,
\end{equation}
which is approximately solved by
\begin{equation}
a_\mu \sim \mathrm{Re} \int \mathrm{d}^{d+1}q \; \mathcal{A}_\mu(q) \frac{\mathrm{e}^{\mathrm{i}q\cdot x - \sqrt{q^2}r}}{r}
\end{equation}
for $q_\mu \mathcal{A}^\mu = 0$, in the long wavelength limit.  Unfortunately, a straightforward check reveals that for certain solutions $\mathcal{L}$ becomes imaginary on this ansatz.  In particular, \begin{equation}
C^2 r^4 a_\mu^\prime a^{\mu \prime} + C^2 r^4 \partial_{[\mu}a_{\nu]} \partial^{[\mu}a^{\nu]} \propto r .  \label{eq:nonlinearfail}
\end{equation}
as $r\rightarrow \infty$, with a coefficient of arbitrary sign.   At $T=0$, the geometry extends to $r=\infty$, and so we conclude that a non-perturbative correction to $a_\mu$ must be made in order for the dynamics to be well-posed: without such a correction, the argument of the square root in (\ref{eq:L41}) is not always positive.\footnote{One might ask whether this breakdown of the nonlinear dynamics could be cured by switching to infalling boundary coordinates, as in the conventional fluid-gravity correspondence \cite{minwalla}.   Such a coordinate choice will not alleviate the problem here, however, because the metric is not dynamical.}  We stress that there is almost certainly a solution to the nonlinear equations of motion for $A_\mu$ which is real-valued;  however, such a solution must necessarily differ non-perturbatively from the zero sound waves in the IR.   As such, we do not expect (\ref{eq:sound}) to be correct -- even qualitatively -- beyond linear response.  We expect that the resulting equations of motion for $J^\mu$ in the boundary theory are nonlocal in space and time.


At finite temperature, we find a cure for the divergence observed in (\ref{eq:nonlinearfail}).   In this limit, the geometry truncates at $r=r_{\mathrm{h}}<\infty$.    Very close to the horizon, the dominant nonlinearities in $\mathcal{L}$ for small but finite amplitudes $a_\mu$ are (at $z=1$) \begin{equation}
\mathcal{L} = \frac{L^{d+2}}{r^{d+2}} \sqrt{1 + C^2 r^4 \left[ -  a_t^{\prime 2}  - \frac{(\partial_t a_i - \partial_i a_t -f a_i^\prime)(\partial_t a_i - \partial_i a_t + f a_i^\prime)}{f} + \partial_{[i}a_{j]} \partial^{[i}a^{j]} \right] } + \mathrm{O}\left(a_\mu^3\right). \label{eq:lastsqrt}
\end{equation}
Upon first glance, there is a term above proportional to $1/f$ that diverges at the horizon.   However, following our discussion near (\ref{eq:mathcalA}), we observe that the infalling boundary conditions are given by \begin{equation}
f a_x^\prime - \partial_t a_x + \partial_x a_t = 0,
\end{equation} 
and this removes the divergenve at the horizon in (\ref{eq:lastsqrt}).   This cancellation generalizes to $z>1$. Hence, we expect that the resulting theory of nonlinear zero sound waves is better behaved.   Still, we do not know if the resulting nonlinear corrections to (\ref{eq:sound}) admit a sensible hydrodynamic interpretation.  In particular, following the same logic as (\ref{eq:nonlinearfail}), we observe that the amplitude of $\partial_\nu a_\mu$ at which the nonlinearities will qualitatively change the nature of the dynamics vanishes as $T\rightarrow 0$.   

The breakdown of nonlinear hydrodynamics in the probe brane models at $T=0$ appears analogous to the fact that the hydrodynamic gradient expansion in the Einstein-Maxwell holographic theory becomes singular as $T\rightarrow 0$, despite ``appearances" that the mean free path is finite even at $T=0$ \cite{surowka}.   
In fact, in the Einstein-Maxwell system, one finds hydrodynamic sound and diffusion poles (for a conventional charged fluid) even at $T=0$ \cite{edalati, davisonsound},  but also observes non-analytic corrections to the gradient expansion at a finite subleading order.   Such non-analytic corrections also arise in probe brane models, at least when $z>1$ \cite{wu}.  It is not clear whether the breakdown of the gradient expansion at subleading orders in derivatives is related to the failure of the perturbative expansion for small perturbations.   More work to resolve these puzzles is warranted.

\section{Conclusion}
At low temperature, the total charge current is a long-lived quantity in holographic probe brane models with $z\le 2$.   Solving the bulk equations of motion in a derivative expansion,  we found an emergent hydrodynamics of the current operator, analogous to the response of weakly disordered fluids with almost conserved momentum.   This hydrodynamics is responsible for both the holographic zero sound mode and the resulting Drude peak.    In particular, this proves that the decay of zero sound at finite temperature is governed by the same decay rate as the Drude peak.   This also leads to a curious example of a quantum field theory with two different ``hydrodynamic" limits:  one at high temperature, and a qualitatively different one at low temperature.

The nonlinear generalization of this novel low temperature hydrodynamics does not appear to be well-behaved.  It is possible that the full, nonlinear equations of motion for the conserved current $J^\mu$ are non-local on the longest length scales.   
It will be interesting to  determine the fate of zero sound at the nonlinear level, and the resulting nonlinear equations of motion for $J^\mu$.

\addcontentsline{toc}{section}{Acknowledgments}
\section*{Acknowledgments}

We thank Andreas Karch, Richard Davison and Sean Hartnoll for useful discussions.  CFC was supported by the Physics/Applied Physics/SLAC Summer Research Program for undergraduates at Stanford University.    AL was supported by the Gordon and Betty Moore Foundation's EPiQS Initiative through Grant GBMF4302.

\begin{appendix}

\end{appendix}

\bibliographystyle{unsrt}
\addcontentsline{toc}{section}{References}
\bibliography{drudebib}

\end{document}